# Paradox of Peroxy Defects and Positive Holes in Rocks

## Part I: Effect of Temperature

**Friedemann T. Freund[#] and Minoru M. Freund[‡*]**

[#] NASA Ames Research Center

MS 245-4; Tel. 650-279-8478; Fax 650-961-7099

Moffett Field, CA 94035-1000

e-mail friedemann.t.freund@nasa.gov

++++++++++++++++++++++++++++++

Department of Physics

San Jose State University

++++++++++++++++++++++++++++++

Senior Scientist, SETI Institute

++++++++++++++++++++++++++++++

[‡] NASA Ames Research Center

Center for Nanotechnology and Advanced Space Materials

Moffett Field, CA 94035-1000

- Deceased

## Abstract

Most non-seismic, non-geodesic pre-earthquake phenomena are believed to be controlled by the stress-activation of peroxy defects in rocks, which release highly mobile electric charges. Though ubiquitous in minerals of igneous and high-grade metamorphic rocks, peroxy defects have been widely overlooked in the past. The charge carriers of interest are positive holes, chemically equivalent to $O^{-}$ in a matrix of $O^{2-}$, physically defect electrons in the $O^{2-}$ sublattice, highly mobile, able to propagate fast and far. $O^{-}$ are oxidized relative to $O^{2-}$. As such $O^{-}$ are not supposed to exist in minerals and rocks that come from deep within the Earth's crust, where the environments are overwhelmingly reduced. The presence of $O^{-}$ appears to contradict thermodynamics. However, there is no conflict. In order to understand how peroxy defects are introduced into common rock-forming minerals, over which temperature window they release positive holes, and how this may be related to pre-earthquake phenomena, we look at peroxy defects in a crystallographically and compositionally well characterized model system: single crystals of nominally high-purity MgO, grown from the melt under highly reducing conditions. During crystallization the MgO crystals incorporate $OH^{-}$ through dissolution of traces of $H_2O$ into the MgO matrix, leading to a solid solution (ss) $Mg_{1-\delta}(OH)_{2\delta}O_{1-2\delta}$, where $\delta \ll 1$. During cooling, the ss leaves thermodynamic equilibrium, turning into a metastable supersaturated solid solution (sss). Using infrared (IR) spectroscopy it is shown that, during further cooling, $OH^{-}$ pairs at $Mg^{2+}$ vacancy sites rearrange their electrons, undergoing a redox conversion, which leads to peroxy anions, $O_2^{2-}$, plus molecular $H_2$. Being diffusively mobile, the $H_2$ molecules can leave the $Mg^{2+}$ vacancy sites, leaving behind cation-deficient $Mg_{1-\delta}O$. During reheating, but in the sss range, the $O_2^{2-}$ break up, releasing positive hole charge carriers, which profoundly affect the electrical conductivity behavior. In igneous mafic and ultramafic rocks, similar changes in the electrical conductivity are observed in the temperature window, where peroxy defects of the type $O_3Si\text{-}OO\text{-}SiO_3$ break up. They release positive holes, which control the electrical conductivity response. Deciphering these processes helps understanding the stress-activation of positive holes along the geotherm.

## Introduction

Monitoring pre-earthquake phenomena and trying to understand them is the only way to make progress toward the elusive goal of forecasting earthquakes. The conventional seismological approach, based on mechanical physics, has not been successful and is mired in the labyrinth of statistical analysis. Combining seismology with recent progress made in geodesy has not improved the situation by much. If ever we want to be successful in forecasting major seismic events, we have to look for non-seismic, non-geodesic pre-earthquake signals. There are many such pre-earthquake indicators [*Freund*, 2013] and much of the information that they carry points to some electrical process or processes, which seem to become activated within the Earth's crust. Since temperatures increase along the geotherm, it is good by starting with changes in electrical properties of minerals and rocks as a function of temperature.

This is Part I of a two-part paper. It addresses a question that is still widely misunderstood or considered controversial in the geoscience community, namely the fact that igneous and high-grade metamorphic rocks can generate highly mobile electronic charge carriers. Normally these charge carriers exist in the rock-forming minerals in an inactive, dormant state. However, when conditions are such that they become activated, they profoundly affect the electrical properties of the rocks.

Part I and Part II deal with the activation of these electronic charge carriers as a function of temperature and stress, respectively. Though our ultimate goal is to understand rocks, which are inherently very complex, Part I starts by looking at the crystallographically simplest and compositionally well-controlled system: magnesium oxide, MgO, single crystals.

Nominally high purity MgO single crystals can be grown from the melt [*Abraham et al.,* 1971]. Their growth conditions are to be characterized as more reducing than any redox condition to which common magmatic or high-grade metamorphic systems in the Earth's crust will be ever exposed. Yet, as we'll show here, these MgO crystals contain peroxy defects, which are thought to be the hallmark of extremely oxidizing conditions. We describe in Part I how peroxy defects are introduced into the matrix of the MgO single crystals despite their origin in a highly reduced environment, how the peroxy defects release highly mobile electronic charge carriers, which profoundly affect the electrical conductivity, and how all this is consistent with thermodynamics. At the end of Part I, the insight gained by studying laboratory-grown MgO crystals is applied to mafic and ultramafic rocks. It may come as a great surprise to many that the minerals of these rocks from natural highly reduced environments also contain peroxy defects despite their overall highly reducing appearance.

Part II goes straight to rocks, addressing the question how peroxy defects are activated by stress, generating by stress the same type of electronic charge carriers as those activated in MgO single crystals by heating. Part II is relevant to earthquake physics and to the question how stress-activated highly mobile electronic charge carriers are linked to the appearance of pre-earthquake phenomena.

The backdrop to this two-part paper is the observation that – paradoxically – any igneous and high-grade metamorphic rock that can be collected at or near the surface of the Earth contains peroxy defects in the matrix of its constituent minerals. Since peroxy defects consist of pairs of oxygen that have converted from their valence state 2– to the oxidized valence state 1–, their presence in those rocks appears to be inconsistent with well-established thermodynamical laws.

Peroxy defects are a hallmark of highly oxidizing conditions. Except as a byproduct of high energy radiation events due, for instance, to radioactive decay processes [*Marfunin*, 1979], $O^-$ should not exist in igneous and high-grade metamorphic minerals and rocks. In fact, those rocks and the minerals they contain have encountered, over the course of their geological history, only reducing, often highly reducing environments. By the time they become exposed to the highly oxidizing conditions at and near the surface the Earth, due to the $O_2$-rich atmosphere, it is kinetically impossible that peroxy defects could have formed inside the matrix of the minerals. Hence, the mere suggestion peroxy defects exist common minerals and rocks appears to be in violation of thermodynamical principles.

However, the presence of peroxy defects is not in violation of thermodynamics. The reason is that peroxy defects are introduced into the matrix of oxide and silicate materials by way of a special, yet apparently ubiquitous solid state reaction that operates outside thermodynamic equilibrium. As will be shown in Part I, taking well-characterized MgO single crystals as a model, peroxy defects are introduced during cooling, in a temperature window when thermodynamic equilibrium no longer applies and the system has entered a metastable state.

It has long been known that peroxy defects exist in oxide materials. A well-studied case is amorphous silica, a-$SiO_2$, where peroxy links $O_3Si\text{-}OO\text{-}SiO_3$ affect the optical properties of fused silica optical fibers [*Azzoni et al.*, 1994; *Fukuchi*, 1996; *Ricci et al.*, 2001]. Another well-studied case is MgO, where the formation of $O_2^{2-}$ anions has been shown to be linked to the incorporation of $H_2O$, in form of $OH^-$ [*Freund and Wengeler*, 1982]. Studying solute $OH^-$ in the MgO matrix has opened the way to understand why peroxy can exist in minerals of igneous and high-grade metamorphic rocks despite their heritage from pervasively reduced crustal and even upper mantle environments.

To discuss what is special about peroxy defects and the positive hole charge carriers, which they produce, we begin with a description as to how peroxy defects are introduced into the matrix of MgO crystals that have taken up small amounts of $H_2O$ component during crystallization from the melt.

We have chosen MgO as a model oxide material because we need to first understand the nature of peroxy defects and the nature of the highly mobile electronic charge carriers that they engender. MgO is the best model material for such a study.

MgO is the most ionic among all alkaline earth oxides with the Coulomb interaction between $Mg^{2+}$ and $O^{2-}$ contributing about 90% to the lattice energy [*Pacchioni et al.*, 1993]. As a main group element Mg has only one oxidation state, 2+. With a band gap around 8 eV, pure MgO should be a textbook example of a near-perfect insulator, assuming that the valence for oxygen is fixed at 2–.

MgO crystals are typically grown under extremely reducing conditions, from the MgO melt produced in an open-pit carbon arc furnace, where an electric arc heats MgO powder to the melting point inside a densely packed MgO powder bed that acts as crucible [*Butler et al.*, 1971]. Using MgO powders of highest chemical purity, up to 99.999%, MgO single crystals can be obtained in purity grades up to 99.99% [*Abraham et al.*, 1971]. However, it is to be noted that, in any purity grade classification, gas/fluid phase components such as $H_2O$ are not counted.

We'll discuss first arc-fusion-grown MgO single crystals and then turn to a rock, dunite, from the highly reduced upper mantle environment and to selected igneous rocks from similarly reduced deep crustal environments. In all cases the question is: do these samples contain peroxy defects?

## Solid Solutions and Supersaturated Solid Solutions

MgO–$H_2O$ is the simplest binary oxide system AO–$H_2O$. **Figure 1** shows the AO-rich side of a binary phase diagram between a high melting point oxide or siliate material AO and the gas/fluid phase component $H_2O$. $T_{melt}$ indicates the melting temperature of pure AO. The presence of $H_2O$ causes $T_{melt}$ to be lowered to the crystallization temperature $T_{cryst}$.

$H_2O$ becomes incorporated into the solid state in form of hydroxyl, $OH^-$, forming a solid solution (ss). The ss field is widest at $T_{cryst}$. Thermodynamics mandates that, with decreasing T, the width of the ss field must decrease. If it were possible to maintain thermodynamic equilibrium, the solute $OH^-$ would be continuously segregated, until the ss field shrinks to zero at 0 K.

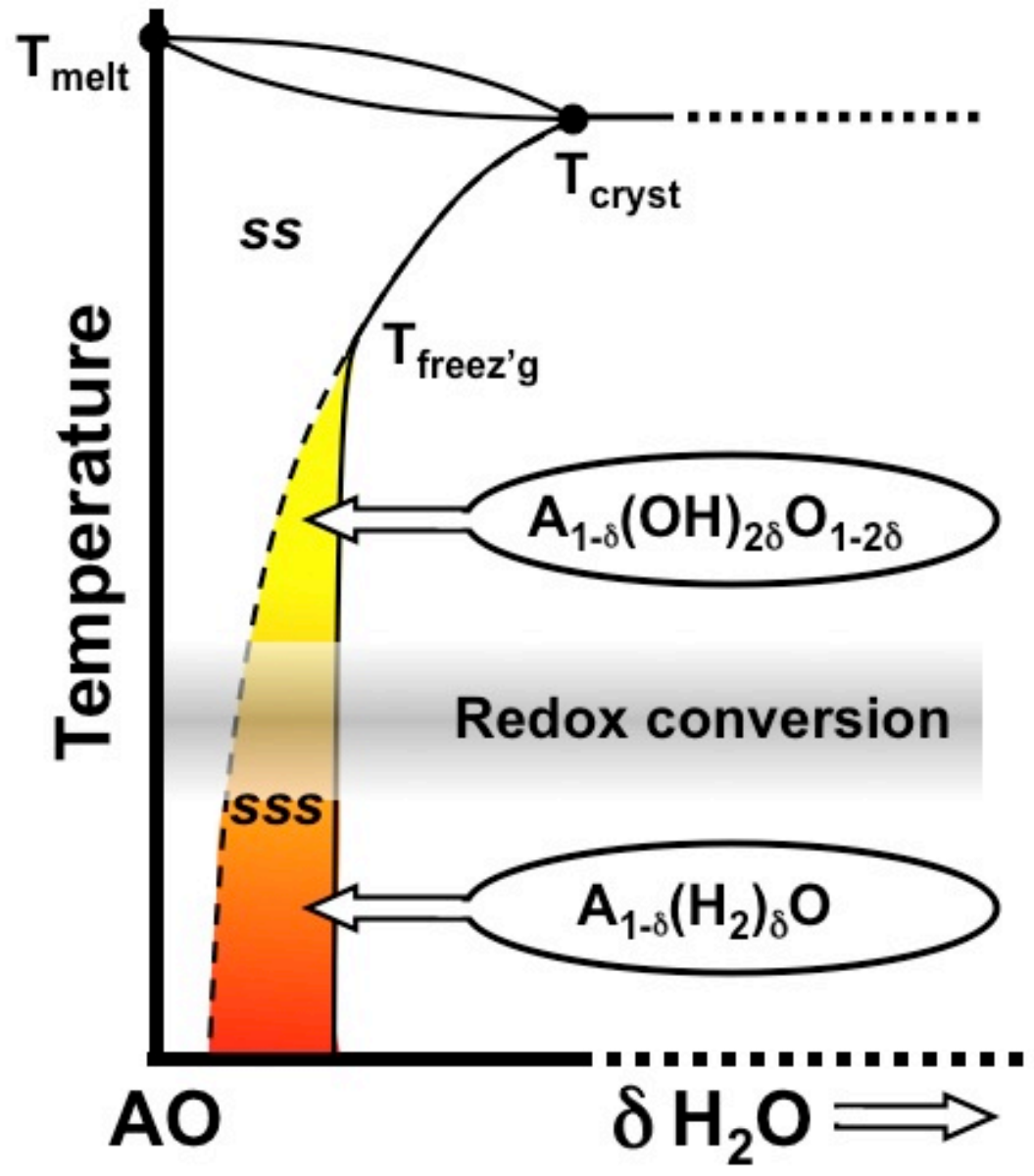

***Figure 1**: AO-side of binary AO-$H_2O$ phase diagram showing the solid solution and supersaturated solid solution fields.*

However, the segregation of solute $OH^-$ is a diffusion-controlled process. Therefore, the system will eventually reach a temperature $T_{freez'g}$, below which segregation becomes so sluggish that the system freezes. When T drops below $T_{freez'g}$, the solid solution leaves thermodynamic equilibrium and turns into a supersaturated solid solution (sss), marked yellow in **Figure 1**. The transition from ss to sss at $T_{freez'g}$ depends on the cooling rate and typically falls into the range around 600°C [*Nesse*, 2000].

MgO crystallizes in the face-centered cubic NaCl-type structure where all $Mg^{2+}$ and $O^{2-}$ are on crystallographically equivalent sites. We write the dissolution of δ $H_2O$ in the MgO matrix as (δ « 1):

$$MgO + \delta\, H_2O \Leftrightarrow Mg_{1-\delta}(OH)_{2\delta}O_{1-2\delta} + \delta\, MgO \qquad [1]$$

Eq. [1] is mass-balanced and states that ss is formed by substituting two $H^+$ for one $Mg^{2+}$. In other words, the introduction of 2δ $OH^-$ causes the introduction of δ $Mg^{2+}$ vacancies.

As long as thermodynamic equilibrium is maintained during cooling, the MgO–$H_2O$ ss continuously adjusts to the narrowing of the ss field. It requires removing $OH^-$ and $Mg^{2+}$ cation vacancies from inside the grain. This process, termed "exsolution", can only be achieved, if excess $OH^-$ and excess $Mg^{2+}$ vacancies diffuse from the interior of the crystal grain to the grain surface or grain boundary. At the interface two $OH^-$ will combine to form $H_2O$ plus $O^{2-}$ with the $O^{2-}$ chargewise compensating for the arrival of the $Mg^{2+}$ vacancies.

The solid state exsolution slows down during cooling, because diffusion slows down exponentially with decreasing T. Regardless how slowly a system is cooled, it will always reach a temperature $T_{freez'g}$, below which the system is considered frozen. Transitioning through $T_{freez'g}$ causes the system to leave thermodynamic equilibrium. Entering the sss field means entering the realm of metastability.

## Solute $OH^-$ in Solid Solution

In order to learn more about solute $OH^-$, we need to know (i) what types of $OH^-$–bearing defects are formed, (ii) where the different $OH^-$ reside in the crystal matrix and (iii) what happens to them during further cooling.

MgO crystallizes in the densely packed, highest symmetry, face-centered cubic (fcc) crystal structure. Because high-purity MgO crystals are transparent over a wide spectral range, we can use infrared (IR) spectroscopy to identify the solute $OH^-$ species that form in the MgO matrix.

The IR absorption bands arising from the O-H stretching modes of different types of $OH^-$ in the MgO matrix, $\nu_{OH}$, have been fully assigned [*Freund and Wengeler*, 1982]. **Figures 2a/b** depict the crystallographic sites where the "impurity" $OH^-$ can reside in the MgO matrix. For the ideal fcc structure, assuming the validity of the substitutional mode as given by eq. [1], there are three types of solute $OH^-$ that can form. The most abundant type should be $OH^-$ pairs at $Mg^{2+}$ vacancy sites.

Using the Kröger-Vinck point defect designation* the $OH^-$ pairs at $Mg^{2+}$ vacancy sites are written as: $[OH^{\bullet}\, V_{Mg}^{''}\, HO^{\bullet}]^x$, where the two $H^+$ of the $OH^-$ pair compensate for the missing 2+ charge of $Mg^{2+}$.

The $OH^-$ pair defects can dissociate according to:

$$[OH^{\bullet}\, V_{Mg}^{''}\, HO^{\bullet}]^x \Leftrightarrow [OH^{\bullet}\, V_{Mg}^{''}\,]' + OH_i^{\bullet} \qquad [2]$$

I II III

where $[OH^{\bullet}\, V_{Mg}^{''}\,]'$ is a single $OH^-$ at an $Mg^{2+}$ vacancy site and $OH_i^{\bullet}$ is an interstitial $OH^-$ with its hydroxyl proton at any regular $O^{2-}$ in the MgO structure. The Roman numerals I, II and III refer to the defects labeled I, II, and III in **Figure 2a**.

Because the $OH^-$ pair defect is neutral with respect to the surrounding MgO matrix, it has the highest probability to form as part of the MgO–$H_2O$ solid solution. Hence, as indicated in the "Expected IR spectrum" in **Figure 2a**, the IR band I associated with two $OH^-$ facing each other across an $Mg^{2+}$ vacancy should give the highest intensity $\nu_{OH}$ band. The $\nu_{OH}$ band II due to a single $OH^-$ at an $Mg^{2+}$

* V stands for vacancy; subscripts identify the site (except for oxygen sites, where subscripts are omitted); superscript prime, dot, and x designate single negative, positive and neutral charges, respectively, double prime and double dot designate double negative and positive charges; subscript i means interstitial; square brackets outline the essential parts of any given point defect [Kröger, F. A. (1964), *The Chemistry of Imperfect Crystals*, North-Holland, Amsterdam.]

vacancy site should have a lower intensity, while the $\nu_{OH}$ band III arising from interstitial $OH^-$ is expected to be excessively broadened due to rapid motion of their $H^+$ from $O^{2-}$ to $O^{2-}$ sites, also known as H-bonding [*Steiner*, 2002].

The "measured IR spectrum" depicted in **Figure 2b** deviates significantly from the "expected IR spectrum". The $\nu_{OH}$ band at 3560 $cm^{-1}$, assigned to $OH^-$ pair defects, has low intensity, while the $\nu_{OH}$ band at 3700 $cm^{-1}$, due to single $OH^-$ at $Mg^{2+}$ vacancy sites, dominates the spectrum. In addition, there is a weak IR band at 4150 $cm^{-1}$. This band arises from a combination of the H–H stretching vibration of molecular $H_2$ in solid matrix with a phonon lattice mode [*Kriegler and Welch*, 1968]. The assignment of this band to $H_2$ has been confirmed through the presence of a $\nu_{HD}$ band in MgO single crystals grown from a melt containing dissolved $D_2O$ [*Freund et al.*, 1982].

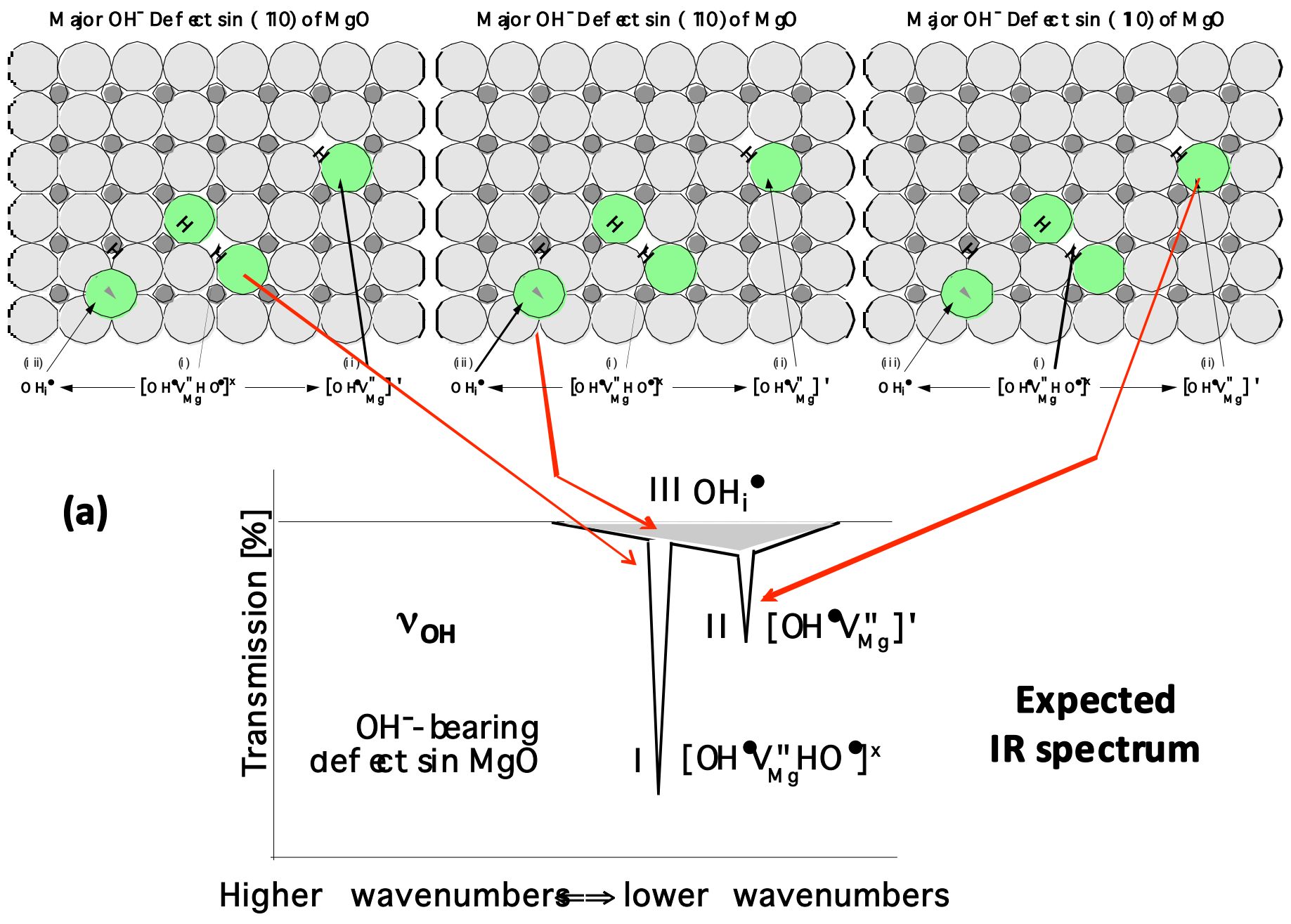

***Figure 2a**: $OH^-$-bearing point defects derived from the dissolution of $H_2O$ in MgO and expected relative intensities of the $\nu_{OH}$ bands due to $OH^-$ on the three dominant $OH^-$–bearing defect sites. Of those three expected $\nu_{OH}$ bands the one labeled I, due to $OH^-$ pairs at $Mg^{2+}$ vacancies, should be most intense [after* [*Freund and Wengeler*, 1982].

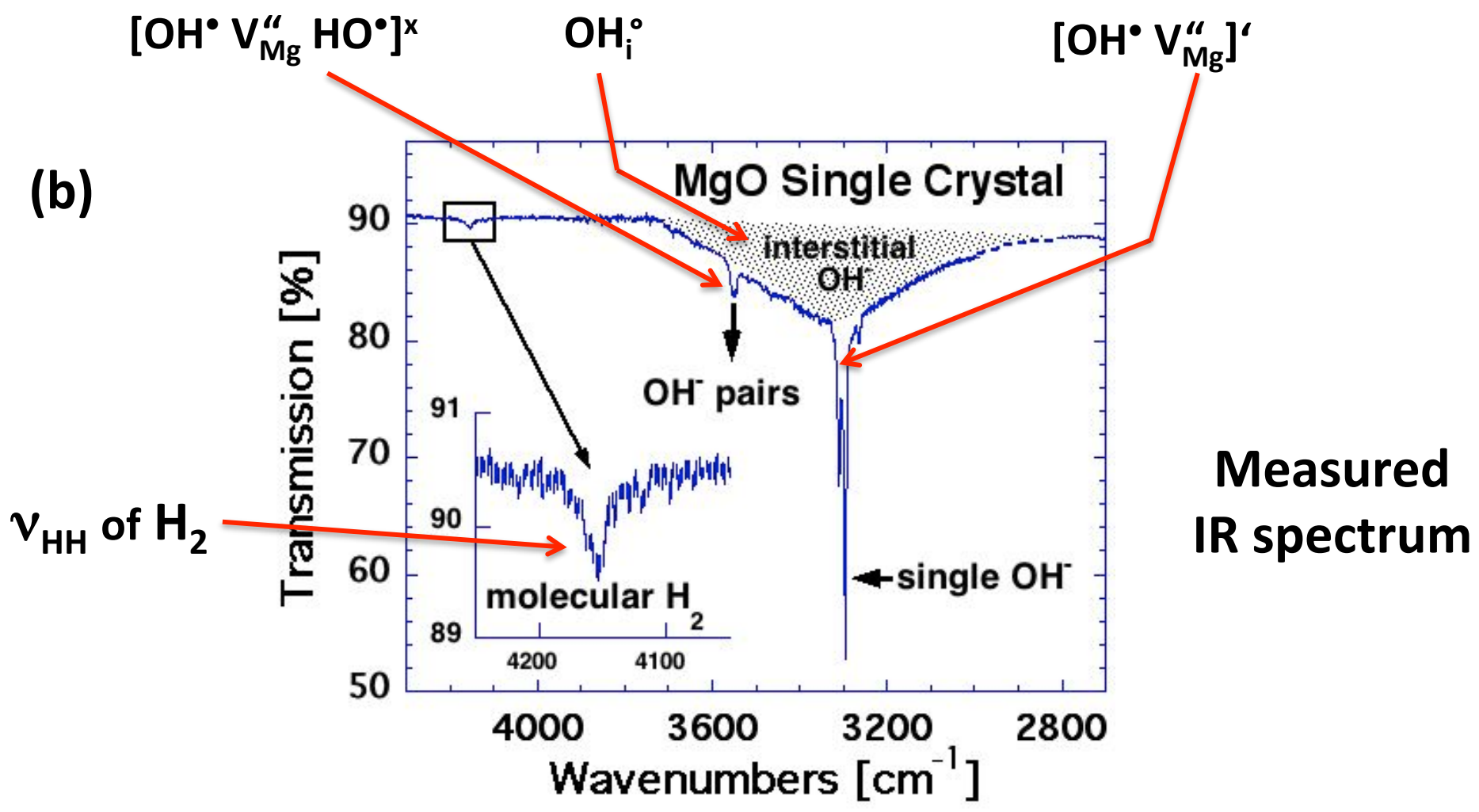

***Figure 2b**: Measured IR spectrum: contrary to expecttion the $\nu_{OH}$ band due to single $OH^-$ at $Mg^{2+}$ vacancy sites is most intense. However, there is also a $\nu_{HH}$ band at 4150 $cm^{-1}$, due to the presence of $H_2$ in the solid MgO matrix (after* [*Freund and Wengeler*, 1982]).

**Redox Conversion of $OH^-$ Pairs to Peroxy plus $H_2$**

The earliest indication that molecular $H_2$ can evolve out of MgO was obtained during a mass spectroscopic study of the gases released during heating of ultrahigh purity $OH^-$–rich MgO produced *in situ* by the thermal decomposition of $Mg(OH)_2$ [*Martens et al.*, 1976]. This study provided evidence for copious amounts of $H_2$ evolving from the nanocrystalline MgO powder, followed by the evolution of atomic O with a sharp on-set temperature of 600°C.

The release of atomic O and its sharp on-set at 600°C are indicative of disproportionation of peroxy:

$$O_2^{2-} => O^{2-} + O \qquad [3]$$

The said mass spectroscopic study also included partially deuteroxylated MgO that produced copious amounts of HD and $D_2$ in addition to $H_2$ [*Freund et al.*, 1982]. This work was substantiated by an IR spectroscopic study of large MgO single crystals, nominally 99.9% and 99.99% pure as well as doped with various 3d-transition metal cations, which were grown under highly reducing conditions from a carbon arc melt [*Freund and Wengeler*, 1982].

Noteworthy is that the $OH^-$–pair defects at $Mg^{2+}$ vacancy sites are expected to be the most abundant $OH^-$–bearing point defects in the MgO–$H_2O$ solid matrix – an expectation that was not confirmed by observation. The $\nu_{OH}$ band due to $OH^-$ pairs at $Mg^{2+}$ vacancy sites was not the most intense but rather weak. The reason is that the $OH^-$ pairs at $Mg^{2+}$ vacancy sites are subject to a redox conversion, in the course of which the hydroxyl oxygens transfer one electron each to their respective protons, reducing the $H^+$ to H, which combine to form $H_2$, while the oxygen anions are oxidized from the valence 2– to the valence 1–, from $O^{2-}$ to $O^-$ which combine to form $O_2^{2-}$. This redox conversion was found to occur during cooling in the 400-600°C window, i.e. over a temperature range when the MgO-$H_2O$ system is already in the metastable sss state, outside thermodynamic equilibrium [*Freund and Wengeler*, 1982]. The temperature window for this redox conversion is marked in gray in **Figure 1**.

Using the Kröger-Vinck point defect designation [*Kröger*, 1964] the redox conversion of the $OH^-$ pairs can be written as:

$$[OH^{\bullet}\, V''_{Mg}\, HO^{\bullet}]^x \Leftrightarrow [{}^{O}_{O}{}^{\bullet\bullet}\,(H_2)''_{Mg}]^x \qquad [4a]$$

Eq. [4a] describes how an $OH^-$ pair at an $Mg^{2+}$ vacancy site in the MgO structure reversibly changes into a peroxy anion, $O_2^{2-}$, and molecular $H_2$.

Since $H_2$ molecules are mobile, even in densely packed MgO, they can diffuse away from the $Mg^{2+}$ vacancy site leaving behind an $Mg^{2+}$ vacancy chargewise compensated by a peroxy anion, $O_2^{2-}$:

$$[OH^{\bullet}\, V''_{Mg}\, HO^{\bullet}]^x \Rightarrow [{}^{O}_{O}{}^{\bullet\bullet}\, V''_{Mg}]^x + (H_2)_i^x \qquad [4b]$$

Eq. [4b] marks the on-set of irreversibility when the $H_2$ molecules formed that the $Mg^{2+}$ vacancy site move away from their "birthplace" and transition into an interstitial site in the MgO matrix.

In the case of nanocrystalline MgO [*Martens et al.*, 1976], the distance from any site in the interior of the nano-size crystallites to the surface is very short, allowing the interstitial $H_2$ molecules to diffuse out of the crystallites and degas. In the case of larger crystals the diffusion will take longer.

$$[OH^{\bullet}\, V''_{Mg}\, HO^{\bullet}]^x \Rightarrow [{}^{O}_{O}{}^{\bullet\bullet}\, V''_{Mg}]^x + (H_2)_{gas} \qquad [4c]$$

Eq. [4c] marks the transition to irreversibility because the $H_2$ molecule has now left the solid state altogether and entered the gas phase. Eq. [4c] describes the formation of cation-deficient $Mg_{1-\delta}O$. If d«1, we may write it as excess-oxygen $MgO_{1+\delta}$.

Eqs. [4a-c] describe a solid state reactions, which have important thermodynamic implications.

First, pairs of redox-neutral $OH^-$ in the MgO matrix convert into $O_2^{2-}$ plus $H_2$, where the peroxy anion consists of two $O^-$, oxidized from the regular $O^{2-}$ state, while the redox-neutral protons of the $OH^-$ have been reduced to the zero-valence state of $H_2$.

Second, the $H_2$ molecules move from their birthplace into interstitial sites, where they are already spatially separated from their oxidized counterpart, the $O_2^{2-}$ anions. Though the overall oxidation state of the system will not change, spatial separation indicated by eq. [4b] will limit the reversibility.

Third, to the extent that $H_2$ are able to diffuse out of the MgO crystals altogether according to eq. [4c], the redox conversion becomes truly irreversible and MgO ends up with excess O in the form of peroxy defects in its crystal structure.

Therefore, MgO crystals, grown from the melt produced under the highly reducing conditions of a carbon arc furnace [*Abraham et al.*, 1971; *Butler et al.*, 1971], acquire peroxy defects, which – under thermodynamic equilibrium conditions – would be the hallmark of highly oxidizing conditions [*Wriedt*, 1987].

As indicated in **Figure 1**, the process described by eqs. [4a/c] does not violate thermodynamics because everything happens in the supersaturated solid solution (sss), i.e. in a temperature window below $T_{freez'g}$, where all major diffusional processes involving cations, anions and cation or anion vacancies have already slowed down to such a degree that the MgO–$H_2O$ system must be considered frozen and in a metastable, non-equilibrium state.

Besides mass spectrometry and IR spectroscopy other techniques have been employed to study the MgO–$H_2O$ sss system. Valuable information was obtained through follow-on single crystal studies of the electrical conductivity [*Kathrein and Freund*, 1983], thermal expansion [*Wengeler and Freund*, 1980], magnetic susceptibility [*Batllo et al.*, 1991], electron spin resonance [*Kathrein et al.*, 1984], dielectric polarization [*Freund et al.*, 1993], refractive index [*Freund et al.*, 1994], and most recently muon spin relaxation [unpublished results]. These additional investigations have provided irrefutable evidence that, despite their provenance from the extremely highly reducing conditions of a carbon arc fusion furnace [*Abraham et al.*, 1971; *Butler et al.*, 1971], the MgO crystals contain peroxy defects and evolve $H_2$ gas [*Freund et al.*, 2002].

## Peroxy Defects

The term "peroxy-" is used in chemistry to designate the bivalent group –OO–. Peroxy entities are well known in organic chemistry, where they form a variety of organic compounds with $R_1$–OO–$R_2$, where $R_1R_2$ = organic ligands [*Klenk and Peter H. Götz*, 2002]. In inorganic chemistry a few stoichiometric peroxy compounds are known, thermodynamically stable at atmospheric $O_2$ pressure or at higher $O_2$ pressures, such as peroxides of alkaline metals $Na_2O_2$, $K_2O_2$, $Rb_2O_2$, and $Cs_2O_2$ [*De la*

*Croix et al.*, 1998] and peroxides of the heavy alkaline earths, in particular of barium, $BaO_2$ [*Jorda and Jondo*, 2001]. These inorganic peroxides are predominantly ionic with $O_2^{2-}$ anions consisting of two $O^-$ engaged in a very short, ~1.5 Å, bond.

In the context of this paper we note that, however the peroxy defects are created, they introduce a semiconductor aspect.

Peroxy anions, $O_2^{2-}$, consist of pairs of oxygen anions that have changed their valence from 2– to 1–. The peroxy bond tightly couples the two $O^-$. In semiconductor parlance, an $O^-$ surrounded by $O^{2-}$ represents a defect electron or "hole" the oxygen anion lattice. A peroxy anion thus represents a self-trapped hole pair. The basic difference between a hole in the oxygen anion sublattice of MgO and a hole in a regular semiconductor material such as silicon, is that, in the latter, holes are introduced by chemically doping Si with an aliovalent impurity such as boron, B. In the case of MgO, the holes are introduced without changing the overall chemical composition but by changing the valence of a few oxygen anions from 2– to 1– and slightly shifting the cation-to-anion ratio.

As long as the two $O^-$ forming a peroxy bond are tightly coupled, their 1– valence is localized in the $O^-$–$O^-$ bond. The presence of $O_2^{2-}$ anions has not detectable effect on the electrical conductivity of the MgO. This changes when the $O^-$–$O^-$ bond breaks.

The $O^-$–$O^-$ break-up can be achieved by heating [*Freund et al.*, 1993]. Thermal dissociation occurs in two clearly identifiable stages. At the outset the $O_2^{2-}$ entity is diamagnetic with the spins associated with each $O^-$ tightly coupled and antiparallel. The break-up starts by loosening of the $O^-$–$O^-$ bond while maintaining spin decoupling and, hence, the diamagnetic state. This is followed by decoupling of the spins, allowing the spins to flip and, hence, to produce a paramagnetic signal [*Batllo et al.*, 1991]. Summarily this dual-step process is shown in eq. [5]:

$$\left[{}^{O}_{O} \,{\bullet\bullet}\, V_{Mg}^{''}\right]^{x} \Leftrightarrow \left[O^{\bullet}\, V_{Mg}^{''}\, O^{\bullet}\right]^{x} \qquad [5]$$

During spin decoupling, the wave functions associated with the two $O^-$ begin to delocalize over many neighboring $O^{2-}$. This causes anomalies in the thermal expansion, refractive index and other material properties. [*Freund et al.*, 1994]. However, as long as the both spins are still bonded to their parent $Mg^{2+}$ vacancy site, the electrical conductivity of the MgO crystals remains weakly affected [*Kathrein and Freund*, 1983].

The break-up completes with the transfer of an electron from an $O^{2-}$ outside the $[O^{\bullet}\, V_{Mg}^{''}\, O^{\bullet}]^{x}$ defect into the decoupled peroxy bond. The result is that the outside $O^{2-}$, which transferred the electron, turns into $O^-$, while the peroxy bond becomes fully dissociated:

$$[O^{\bullet}\, V_{Mg}''\; O^{\bullet}]^{x} \;\Leftrightarrow\; [O^{\bullet}\, V_{Mg}''\;]' \;+\; O^{\bullet} \qquad [6]$$

The first defect on the right had side of eq. [6], an $Mg^{2+}$ vacancy associated with one $O^{-}$, known as $V^{-}$ center, a paramagnetic defect. It has been intensely studied by electron spin resonance spectroscopy [*Henderson and Wertz*, 1977; *Marfunin*, 1979].

The outside $O^{2-}$, which donated an electron, turns into $O^{-}$ as shown on the right had side of eq. [6]. No longer bound to the $Mg^{2+}$ vacancy site, this $O^{-}$ turns into a highly mobile, charge carrier, a defect electron in the $O^{2-}$ sublattice, for which the name "positive hole" has been proposed [*Griscom*, 1990].

The peroxy dissociation according to eq. [6] is written as a reversible reaction to indicate that, under the appropriate conditions, for instance during cooling, positive holes recombine with the $V^{-}$ centers, returning to the peroxy state with two tightly coupled $O^{-}$.

The forward reaction described by eq. [6] – dissociation – is a second-order phase transition. It starts around 430°C and reaches completion around 600-650°C. The back reaction – recombination – tends to be sluggish, taking minutes to proceed during cooling, hours at room temperature.

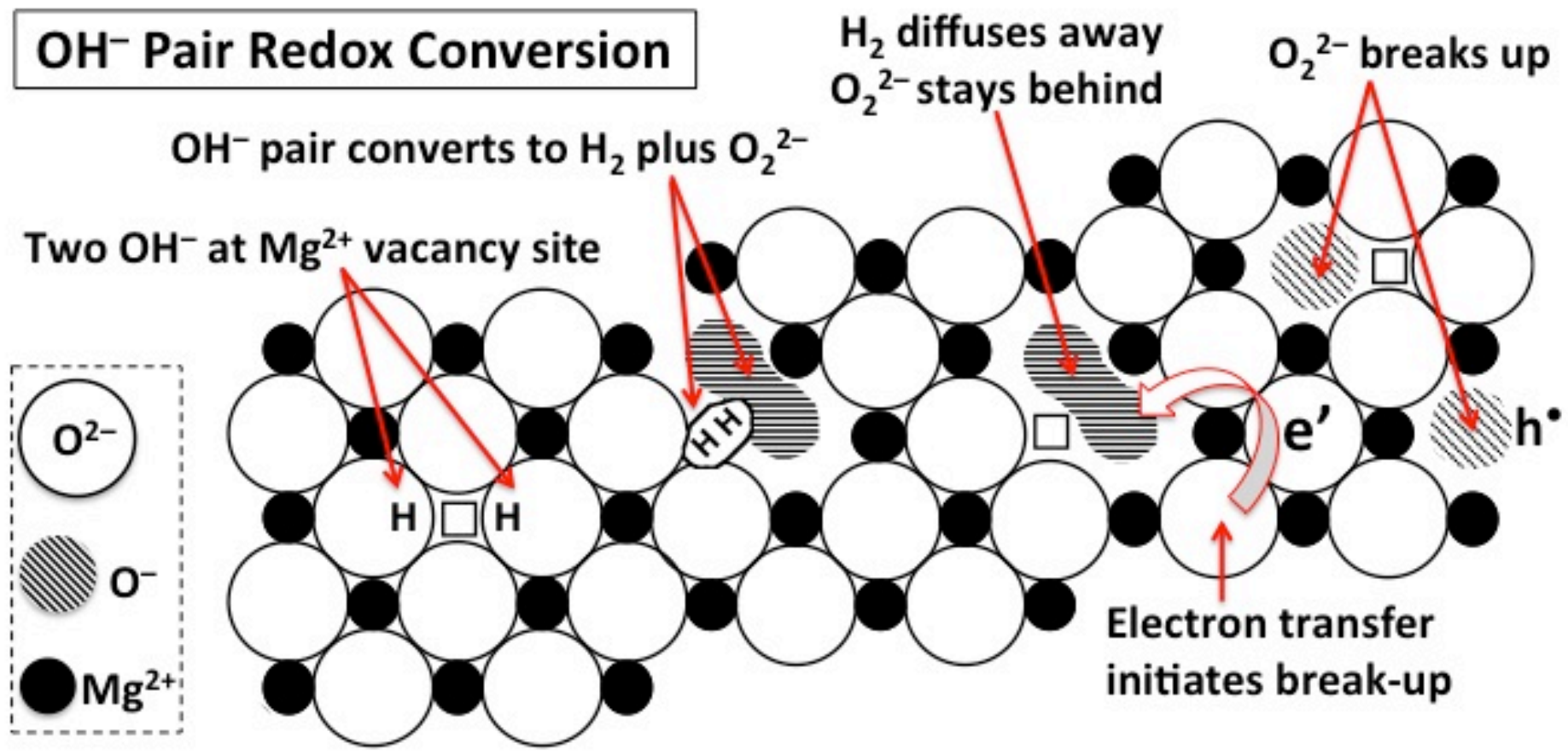

***Figure 3**: Depiction of the sequence of steps at the $OH^{-}$ pair defect site in MgO, the $OH^{-}$ pair redox conversion to $O_2^{2-}$ plus $H_2$, the loss of $H_2$ from the site, and the break-up of the peroxy anion, $O_2^{2-}$, into a bound $O^{-}$ plus an unbound $O^{-}$.*

**Figure 3** summarizes the multi-step process as just described, starting from the left with an $OH^{-}$ pair and its redox conversion to $O_2^{2-}$ plus $H_2$ according to eq. [4a]. Next comes the stage when the $H_2$ has diffuses away, leaving behind the peroxy at the $Mg^{2+}$ vacancy site according to eqs. [4b/c]. The step describing the decoupling of the spins is omitted. Instead, we show on the right the break-up of the peroxy anion, leaving one $O^{-}$ at the $Mg^{2+}$ vacancy site, while the other $O^{-}$ becomes unbound, turning into a positive hole charge carrier, $h^{\bullet}$, according to eq. [6].

## Positive Holes

Positive holes, symbolized by $h^{\bullet}$, are defect electrons in the $O^{2-}$ anion sublattice [*Griscom*, 1990]. In semiconductor parlance, positive holes are virtual particles associated with electronic wave functions that describe a missing electron in some otherwise fully occupied energy levels at the upper edge of the valence band.

**Figure 4** illustrates the start and end of the peroxy break-up. The upper left shows an intact peroxy defect, characterized in bright red by a dumbbell $O_2^{2-}$ anion next to an $Mg^{2+}$ vacancy. The upper right shows a dissociated peroxy defect. The red color of the $O^{2-}$ surrounding the $Mg^{2+}$ vacancy in the plane of the drawing, indicates that the $O^{-}$ state delocalizes over the neighboring $O^{2-}$. The unbound $O^{-}$, i.e. positive hole, is represented by shades of red and rose spread over many $O^{2-}$, indicating that the wave function associated with $h^{\bullet}$ is highly delocalized.

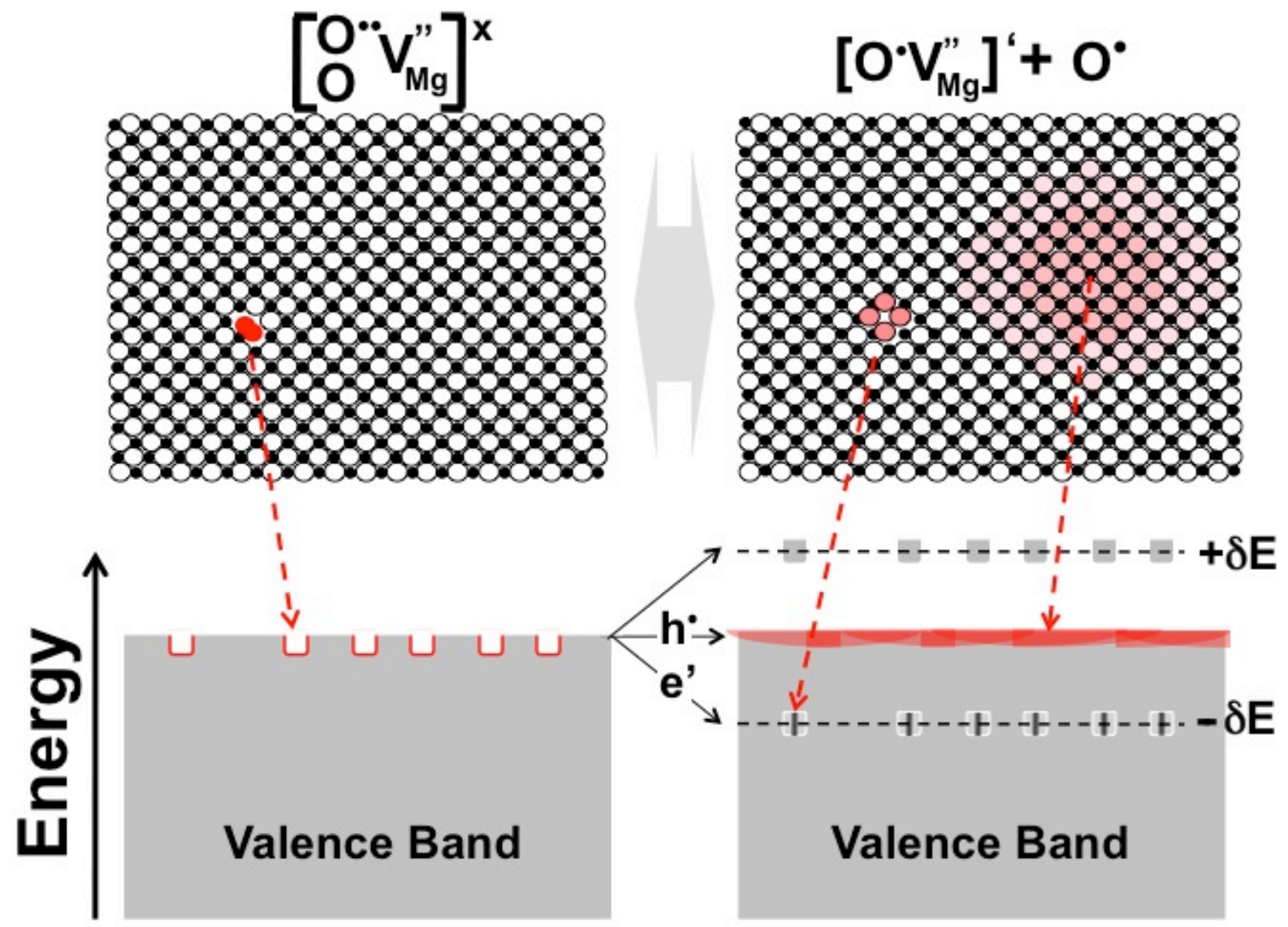

***Figure 4**: Top left: Projection of the (100) plane of MgO with a peroxy anion, $O_2^{2-}$, at an $Mg^{2+}$vacancy site; Bottom left: Dip in the energy surface at the edge of the valence band due to the presence of peroxy anions, $O_2^{2-}$; Top right: Dissociated state of the peroxy defect with one $O^{-}$ at the $Mg^{2+}$ vacancy site, delocalized over its immediate neighbors, and the highly delocalized wave function of the $h^{\bullet}$; Bottom row: Splitting of the energy levels during dissociation of the peroxy defect to generate positive holes and electrons in shallow traps (see text).*

The lower part of **Figure 4** depicts energy levels that arise from the presence of peroxy defects and positive holes. The edge of the valence band is made of energy levels that derive mostly from O 2sp-symmetry atomic orbitals of $O^{2-}$. The highest occupied level, associated with the $O^{2-}$, is antibonding and of σ* symmetry. In the peroxy anion the σ* level is empty, leaving the non-bonding $\pi^{nb}$ as the highest occupied level [*Marfunin*, 1979]. Peroxy defects can thus be represented by a local dip in the energy surface of the valence band as schematically depicted in the lower left side of **Figure 4**.

When the peroxy bond breaks, the non-bonding, $\pi^{nb}$ energy level splits as indicated in the lower right of **Figure 4**. The transfer of an electron into the broken peroxy bond creates new energy levels below the edge of the valence band, shifted downward by -δE. The electron trapped by the broken peroxy bonds will occupy this level. Due to symmetry considerations, a corresponding energy level must be created in the forbidden band gap, shifted upward by +δE. The delocalized wave functions associated with positive holes $h^{\bullet}$ is indicated by the red smear at the upper edge of the valence band.

## DC Conductivity of High-Purity MgO Single Crystals

The insight gained so far can be applied to direct current (dc) electrical conductivity measurements of MgO single crystals over the temperature range, in which positive hole charge carriers can exist.

Well-annealed MgO single crystals are nearly perfect insulators. Using a guard electrode to remove contributions from surface conductivity, their room temperature conductivity was «$10^{-16}$ [$\Omega^{-1}cm^{-1}$] [*Kathrein and Freund*, 1983]. Upon heating the MgO crystals display a complex response, which allows us to correlate the conductivity behavior with the activation of positive hole charge carriers as described in the preceding section.

For thermally activated conductivity due to charge carrier i with activation energy $E_i$ we have:

$$\sigma = \sigma_i \exp [-E_i/kT] \qquad [7]$$

where $\sigma_i$ is the pre-exponential factor for the charge carrier i. Plotting log σ versus 1/T eq. [7] gives a straight line, an Arrhenius plot, with the slope $E_i$. If several charge carriers 1, 2, … are involved, endowed with activation energies $E_1$, $E_2$ …, we have:

$$\sigma = \sigma_1 \exp [-E_1/kT] + \sigma_2 \exp [-E_2/kT] + \ldots \qquad [8]$$

Eq. [8] will produce a plot with two or more straight Arrhenius sections corresponding to $E_1$, $E_2$…

The total conductivity is the sum of the contributions by each type of charge carrier:

$$\sigma = \Sigma\sigma_i = \Sigma\, n_i\, z_i\, \mu_i \qquad [9]$$

where $n_i$ are the number densities, $z_i$ the charges, and $\mu_i$ the mobilities.

**Figures 5a-d** shows the dc electrical conductivity response of a nominally high purity 4N-MgO single crystal as a function for temperature under clean conditions in an ultra-dry argon atmosphere, using 2-electrode, 3-electrode- and 4-electrode configurations [*Kathrein and Freund*, 1983]. Here we specifically discuss a series of experiments where an MgO single crystal was repeatedly heated and cooled at 10°C min$^{-1}$.

**(a)**

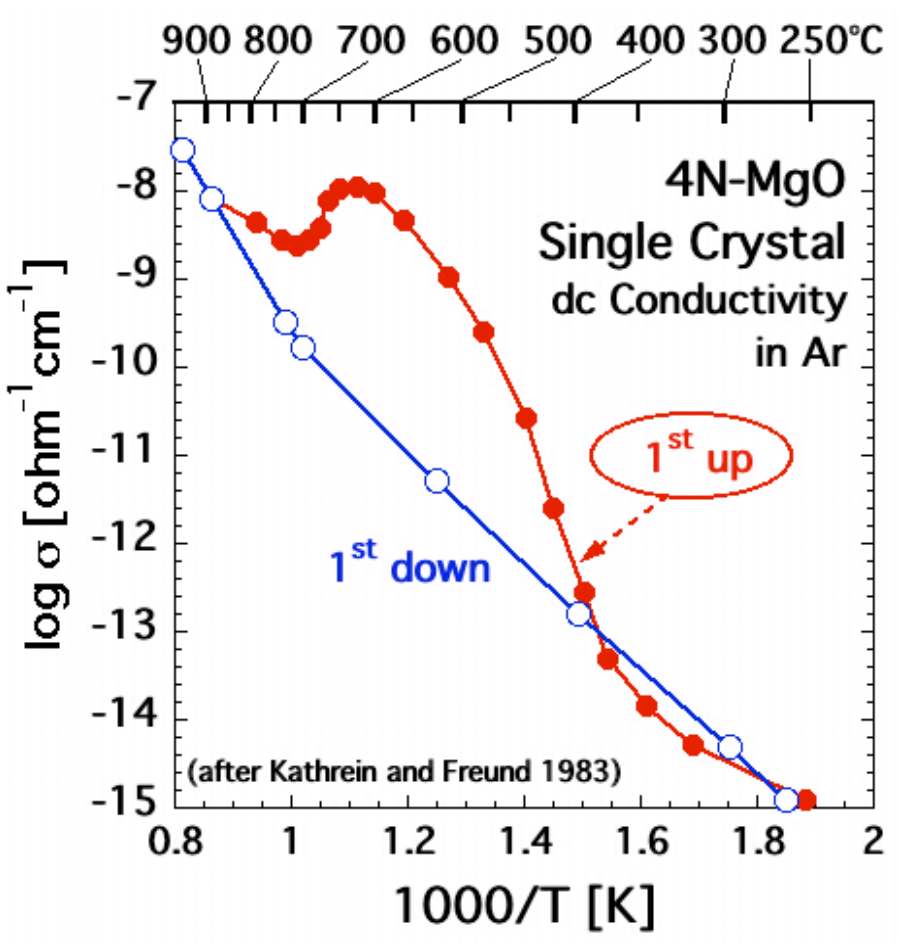

**(b)**

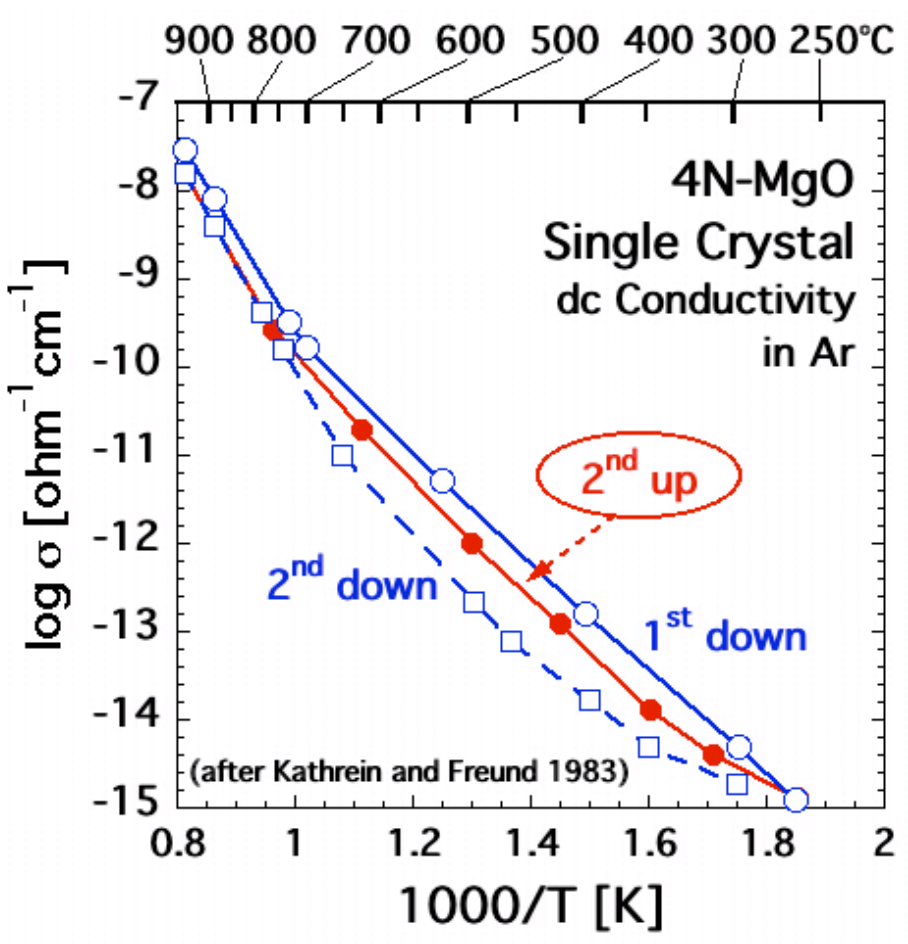

**(c)**

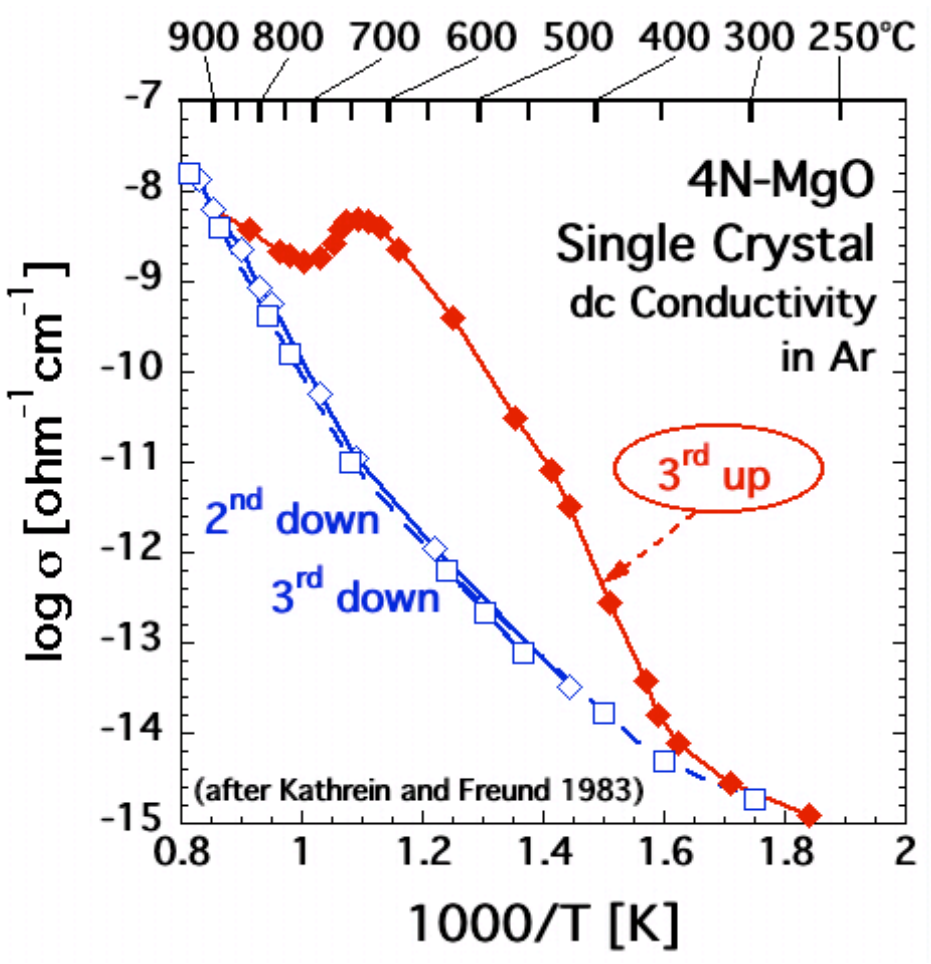

***Figure 5:** Sequence of dc conductivity curves of a highest purity grade MgO single crystal under the stringent clean conditions. Panel **(a)** shows an anomalous increase in conductivity during the first heating, which is not observed, when the crystal is reheated immediately after cooling as shown in Panel **(b)**. However, if the crystal is annealed over night, the conductivity anomaly during heating reappears as indicated in Panel **(c)** [after [Kathrein and Freund, 1983]].*

When the MgO crystal was well-annealed, first time heating resulted in the conductivity curve labeled "1st up" in **Figure 5a**. It shows the conductivity rising sharply between 350–400°C, reaching a maximum around 600°C. Above 650°C. the conductivity decreases until 700°C and then increases again. Around 850-900°C the conductivity joins an Arrhenius straight section where the conductivity versus temperature curve becomes fully reversible with an activation energy of 2.4 eV [*Lempicki*, 1953]. In this high temperature range the electrical conductivity is known to be controlled by $Mg^{2+}$ diffusion via an $Mg^{2+}$ vacancy hopping mechanism [*Sempolinski and Kingery*, 1980], possibly augmented by $O^{2-}$ diffusion [*Wuensch et al.*, 1973]. During cooling, labeled as "1st down" in **Figure 5a**, the conductivity initially follows the reversible high temperature curve with an activation energy of 2.4 eV and then transitions to a lesser slope, corresponding to a 1 eV activation energy, indicating a different conductivity mechanism.

In **Figure 5b** we replot the "1st down" curve together with the "2nd up" curve obtained by reheating the MgO crystal immediately after it had cooled to about 250°C. The "2nd up" curve differs markedly from the "1st up" curve: it stays close to the cooling curve "1st down". At about 700°C the "2nd up" curve displays the same 2.4 eV slope as before. The "2nd down" cooling curve follows the same trend including consistently lower overall conductivity values along its path.

This pattern of dc conductivity as a function of heating and cooling is reproducible [*Kathrein and Freund*, 1983]. It is observed not only with MgO crystals of nominal 99.9% and 99.99% purity grades but it is also seen in essentially the same magnitude with MgO crystals doped at different concentration levels with 3d-transition metal cations such as $Mn^{2+}$, $Fe^{2+}$, $Co^{2+}$, or $Ni^{2+}$, indicating that the conductivity is not controlled by metal cation impurities nor to any other chemical impurity. The only alternative for the anomalous conductivity response is that it is caused by the presence of peroxy defects and the positive hole charge carriers, which the peroxy defects release.

For the plot in **Figure 5c** the MgO crystal was reheated after having been annealed overnight. The "3rd up" curve exhibits the same anomalous conductivity increase above 350-400°C as during the initial heating, "1st up". The cooling curve, "3rd down", follows the trend observed during the prior two cooling cycles.

Key to understanding this complex temperature-time-dependent conductivity behavior are eq. [5] describing the decoupling of $O^{-}$–$O^{-}$ bond and eq. [6] describing the release of mobile positive hole charge carriers by way of a second order phase transition over the temperature window from 430°C to about 600°C [*Kathrein and Freund*, 1983].

According to this interpretation, the conductivity increase during heating as shown in **Figures 5a** and **5c** signals the activation of an increasing number of positive hole charge carriers. In this case eq. [8] applies, which states that two reactions take place simultaneously with two activation energies $E_1$ and $E_2$ describing respectively the break-up of the peroxy defects and the positive hole conduction proper.

Combining eqs. [8] and [9] gives

$$\sigma = n_1 z_1 \mu_1 \exp[-E_1/kT] + n_2 z_2 \mu_2 \exp[-E_2/kT] \qquad [10]$$

where $E_1$ is the activation energy needed to break up the $O^-$–$O^-$ bonds leading to an increase the number $n_1$ of positive holes $h^\bullet$, and $E_2$ is the activation energy that controls the conduction by $h^\bullet$.

To test the validity of this approach we designed an experiment to separate the two mechanisms depicted in **Figures 6a/b** using an MgO single crystal of 99.99% nominal purity with a guard electrode to minimize any residual surface conductivity contribution. We heated the crystal in dry ultrapure Ar to 250°C to bake out potential surface contamination. After annealing for over 48 hrs, we applied a 20°C/min heating-cooling program, which shifted all temperatures relative to the earlier 10°C/min heating-cooling program. The temperatures given below refer to readings off **Figures 6a/b**.

During first heating we raised the temperature to 550°C, reversed to cooling. After reaching 330°C we switched to reheating, raising the temperature to 580°C, followed by cooling to 370°C. We then reheated it to 610°C and switched again to cooling. After the sample had cooled to 240°C, we once more reheated it.

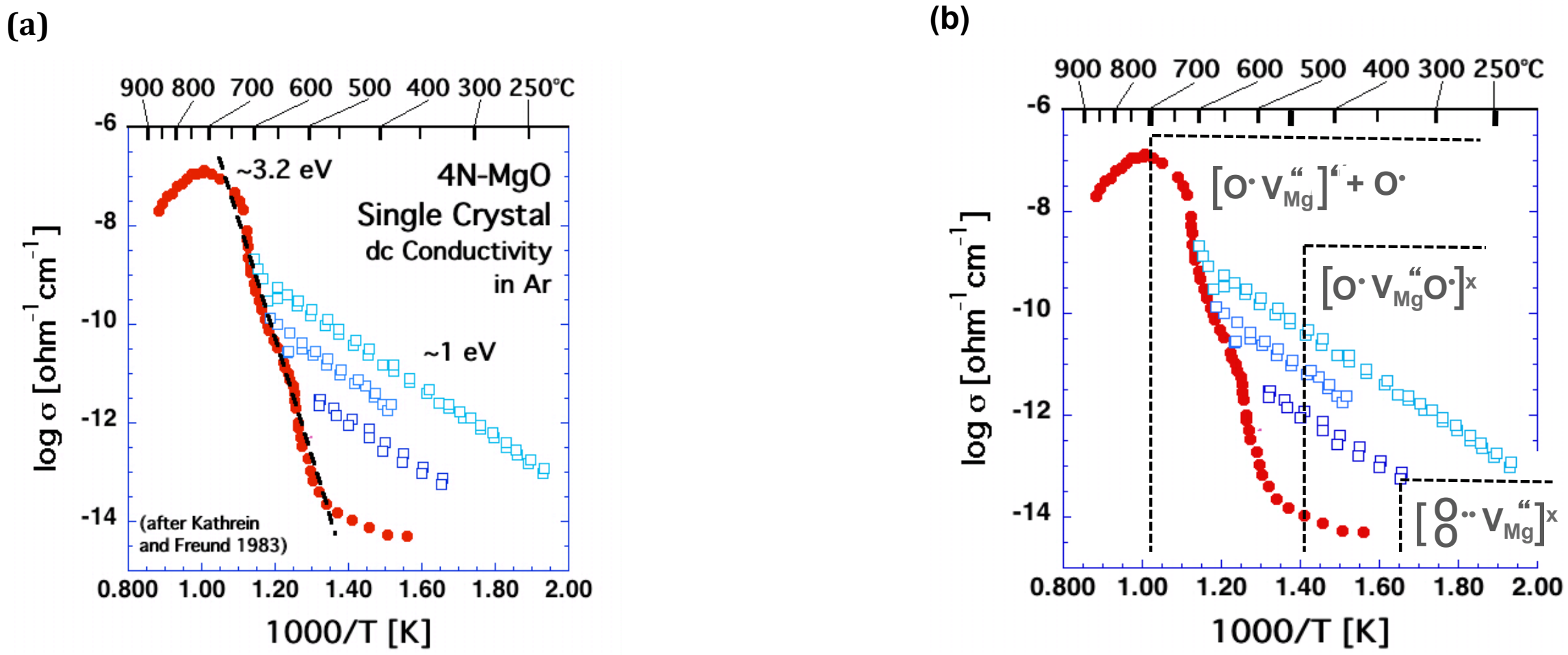

***Figure 6a/b****: Arrhenius plot of the data obtained during a specially designed heating-cooling program to separate conductivity contributions arising from the progressive break-up of peroxy defects (solid red circles) and from positive holes (open squares).*

The conductivity pattern resulting from this heating-cooling program depicted in **Figure 6a/b** can be broken down into parts of interest here:

(i) a steeply rising section (red filled circles) that can be approximated by an Arrhenius straight line over the 480–650°C range (dashed black line) with an activation energy of 3.2 eV;

(ii) three parallel cooling-reheating straight sections (open squares in shades of blue), which suggest a distinct mechanism, endowed with an activation energy of about 1 eV.

**Figure 6b** shows the same experimental data with the point defects of interest inserted from the tightly bounds peroxy defect at right, the spin-decoupled peroxy defect in the middle, and the dissociated peroxy defect at left. The temperature ranges over which these stages exist are indicated.

If (i) is due to the break-up of the $O^{-}–O^{-}$ bonds according to eq. [6] and generation of $h^{\bullet}$ charge carriers, then eq. [8] applies. In this case we can rewrite eq. [8] as:

$$\sigma_i = \sigma_0 \exp\left[-(E_1+E_2)/kT\right] \qquad [11]$$

where $E_1$ refers to the energy needed to break the peroxy bond, creating increasing numbers of $h^{\bullet}$ charge carriers, while $E_2$ refers to the activation energy for $h^{\bullet}$ diffusion.

If $(E_1 + E_2) = 3.2$ eV and $E_2 = 1$ eV, then $E_1 = 2.2$ eV. Thus we note that the energy needed to break the peroxy bond in the MgO matrix is about 2.2 eV.

The activation energy for $h^{\bullet}$ diffusion in the MgO matrix is about 1 eV and the entire electrical conductivity is controlled by $h^{\bullet}$ as the only mobile charge carriers.

## Discussion

Conventionally, earthquake and pre-earthquake research as conducted within seismology takes a mechanical approach, focusing on mechanical aspects of stress build-up, rock deformation and rock rupture. However, there is mounting evidence that, prior to major earthquakes, non-seismic effects become observable, from which valuable information about pre-earthquake conditions can be derived [*Freund*, 2013]. Many of the non-seismic pre-earthquake phenomena point to electrical processes that take place in the rocks of the Earth's crust during the build-up of stress, expressing themselves at the Earth's surface and above. It is therefore of paramount importance to understand the electrical properties of rocks throughout the Earth's crust and how they change as a function of the two main parameters: temperature and stress. Part I of this two part paper is dedicated to temperature.

There is, of course, a large body of literature describing the electrical conductivity of minerals and rocks as a function of temperature. In most studies the focus was on achieving thermodynamic equilibrium conditions. This limits their usefulness with respect to processes that take place in the Earth's crust, where – as suggested by **Figure 1** – the prevailing conditions lie mostly outside thermodynamic non-equilibrium and where metastability is much more the rule than an exception. The number of prior studies that address metastable equilibrium conditions is small, ostensively because such measurements are more difficult to do and to interpret.

Because of these difficulties, we have presented here a broad study of the MgO–$H_2O$ system aimed at elucidating the intricacies of electrical conductivity under conditions of thermodynamic metastability. We have introduced MgO crystals, grown from the melt, as a model for the AO–$H_2O$ phase diagram in **Figure 1,** where AO can be any oxide material or any silicate mineral. The study of MgO has allowed us to unravel the complex solid state processes that result from the incorporation of $H_2O$, which is a universal process during crystallization from any fluid-laden melt or magmas or during recrystallization in any fluid-rich high-grade metamorphic crustal environment. MgO has allowed us to discover a fundamental reaction that takes place in the supersaturated solid solution (sss) state, outside thermodynamic equilibrium, namely the redox conversion (RC) of pairs of solute $OH^-$ to peroxy defects plus $H_2$. This RC reaction is not limited to MgO but occurs across a wide range of minerals and rocks. Focusing on MgO has allowed us to characterize peroxy defects and to understand them as the source of highly mobile positive hole charge cariers, which have a dramatic effect on the electrical conductivity behavior.

Taking advantage of the favorable properties specific to MgO we have documented the RC of $OH^-$ pairs yielding peroxy, $O_2^{2-}$, plus molecular $H_2$. Though first discovered in MgO more than 30 years ago [*Freund and Wengeler*, 1982; *Martens and Freund*, 1976] and further described in follow-up publications for MgO and a few selected natural minerals [*Freund*, 1985; 1987], this RC appears to take place in many mineral-$H_2O$ systems of crustal interest, i.e. under the non-equilibrium, metastable conditions as we presented here. Unfortunately, though this RC is probably ubiquitous across much of the Earth's crust, it has been overlooked or summarily ignored by the geoscience community.

The fundamental insight, which the MgO–$H_2O$ system has provided, is (i) that the valence of oxygen anions in an oxide matrix can change from $O^{2-}$ to $O^-$outside thermodynamic equilibrium, (ii) that the presence of peroxy and $O^-$ has a profound effect on the electrical conductivity in a critical temperature window, up to about 600°C, where metastable conditions prevail, and (iii) that the repeatability of the measured conductivity vs temperature response is consistent with metastable equilibria.

If the presence of $O^-$, i.e. $h^{\bullet}$ charge carriers, has a profound effect on the electrical conductivity response of nominally high purity MgO single crystals forming a dilute AO–$H_2O$ solid solution, we may expect the same effect to be observable in other AO–$H_2O$ systems, where AO are silicate minerals, crystallographically and compositionally more complex than MgO. We may thus expect to see telltale signs of $h^{\bullet}$ charge carriers in laboratory experiment with crustal minerals and crustal rocks and even with minerals and rocks from the Earth's upper mantle which cooled down during their transport to the Earth's surface and necessarily passed through the RC temperature window.

Among the diagnostically most distinct signs of $h^\bullet$ charge carriers are (i) a steep, anomalous increase in electrical conductivity during heating through the temperature window over which peroxy defects dissociate injecting highly mobile $h^\bullet$ into the valence band, and (ii) a seemingly characteristic 1 eV activation energy observed during cooling and/or immediate reheating of samples after cooling.

There are very few, if any, reports in the geoscience literature that describe the steep increase in electrical conductivity during initial heating. The reason is that, if researchers observed such a seemingly anomalous conductivity behavior, they were inclined to assign it to some kind of surface contamination. Such an interpretation, namely surface contamination, indeed appears supported by the observation that, during relatively rapid cooling and relatively rapid reheating, the anomaly have disappear as was demonstrated for MgO in **Figures 5a/b**. The fact that the anomaly reappears upon leaving a given sample at room temperature for an extended period of time, such as overnight as shown in **Figure 5c**, would be dismissed on account of assumed re-contamination. Obviously, so long as the anomalous conductivity behavior is believed to be due to adventitious contamination, it would hardly ever be included in any serious publication.

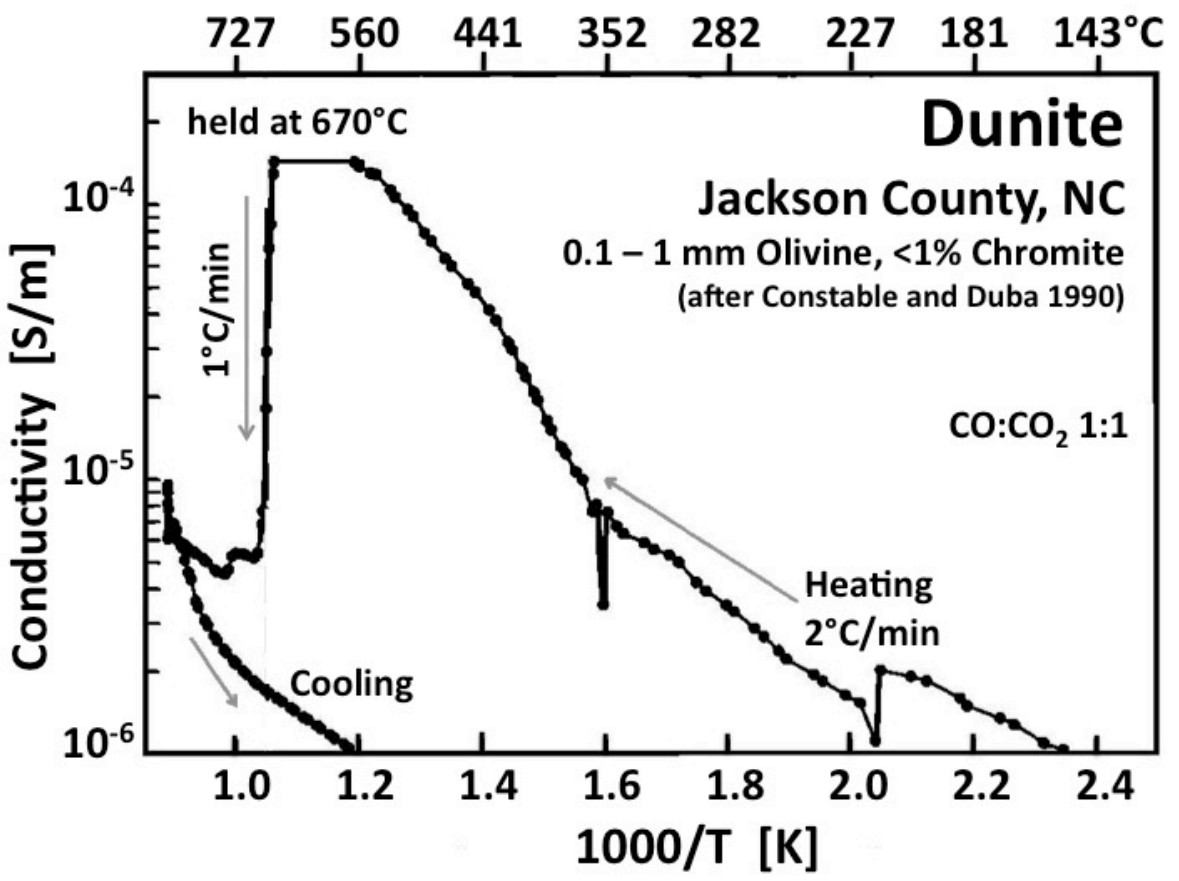

***Figure** 7: Electrical conductivity of peridotite, 99% olivine, heated and cooled in a 1:1 CO:$CO_2$ mixture, exhibiting an anomalously high conductivity region interpreted to be caused by vapor-phase deposited surface carbon and the beginning of a cooling curve, where the slope is consistent with a 1 eV activation energy (after [Duba and Constable, 1993]).*

There is one notable exception, where the initial heating curve has been published for a dunite, an olivine-rich peridotite of upper mantle origin [*Constable and Duba*, 1990]. The Arrhenius plot in **Figure 7** shows a conductivity increase during initial heating in a 1:1 CO:$CO_2$ atmosphere at 1 bar. The conductivity values, derived from data recorded at 2°C/min, were found to increase moderately from about 150°C to ~350°C. Above 350°C the slope becomes steeper up to about 430°C, then slightly less steep up to 560°C, strikingly similar to the conductivity vs temperature behavior of the highest purity MgO single crystal as depicted in **Figure 5a/c**. No data are shown for the 560-670°C interval. At 670°C, the temperature was held constant for an unspecified time, leading to a significant decrease in conductivity. When heating was resumed, the conductivity increased, transitioning above 750°C into a high temperature branch similar to that recorded with MgO single crystals. During cooling the conductivity curve changes around 600°C to an Arrhenius straight section with a ~1 eV activation energy, again similar to the cooling curve observed for MgO.

The interpretation offered by Constable and Duba for the anomalously high conductivity is based on the assumption that, during heating in a 1:1 CO:$CO_2$ mixture, a thin carbon film would deposit onto the sample surface, providing a low resistivity pathway [*Constable and Duba*, 1990]. According to thermodynamic calculations, the 1:1 CO:$CO_2$ mixture would leave the stability field of graphite at 670°C, thereby ending the carbon deposition and even causing the existing carbon film to "burn off". No control experiments are reported such as measuring the electrical conductivity during repetitive heating-cooling-reheating cycles in the same 1:1 CO:$CO_2$ mixture or different gas mixtures to explore the presumed deposition of carbon film with respect to the CO-C phase boundary. We submit that the carbon film formation has not been independently verified and that other causes for the anomalous conductivity behavior of the dunite rock must be considered, for instance conduction by $h^\bullet$.

There are also very few published data on the 1 eV activation energy mechanism associated in MgO with $h^\bullet$ conduction in the lower temperature range. Out of the many papers that address the question of the electrical conductivity of MgO only one paper mentions specifically the 1 eV activation energy observed during cooling below 700°C but offers no explanation [*Lewis and Wright*, 1968 ]. Some other papers on MgO include figures with short sections of a 1 eV Arrhenius plot but remain silent as to a likely conduction mechanism.

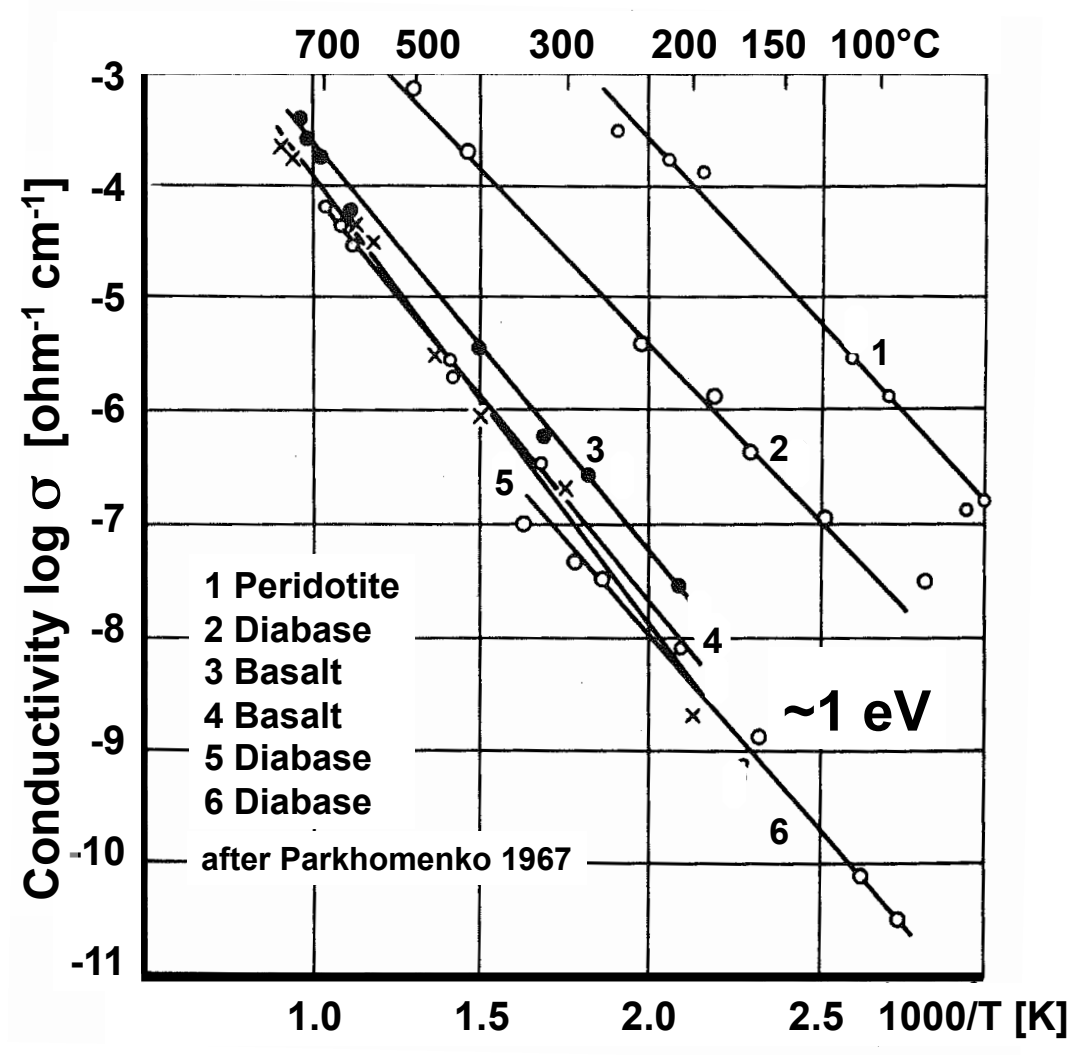

*Figure 8: Electrical conductivity of mafic and ultramafic rocks after [Parkhomenko, 1967; Parkhomenko, 1982].*

The 1 eV activation energy has been well documented for mafic and ultramafic igneous rocks as demonstrated in **Figure 8** *[Parkhomenko, 1967; Parkhomenko, 1982]*. The data were obtained with heating-cooling protocols similar to that used to collect our MgO single crystal data depicted in **Figure 5b**. Over the temperature range from below 100°C to over 700°C, different rocks with very different absolute conductivities exhibit the same 1 eV activation energy. No explanation has been offered in this and other publications except by proposing that the 1 eV conduction may occur along grain boundaries or through intergranular carbon or water films [*Constable and Duba*, 1990; *Roberts et al.*, 1999; *Shankland et al.*, 1997; *Yoshino and Noritake*, 2011]. The possibility of bulk conductivity, through the interior of the crystals, has not been considered.

With polycrystalline materials such as rocks it is inherently difficult, if not possible, to unambiguously distinguish between bulk and surface or grain boundary conduction. In our MgO single crystal study the application of a guard electrode allowed us to ascertain that the 1 eV mechanism was due to charge carriers flowing through the bulk – not along the surface. The fact that, as depicted in **Figure 8**, different rocks display the same 1 eV despite very different absolute conductivity values, makes it unlikely that the current flow be controlled by grain boundaries or intergranular water or carbon films. Instead we propose that there is only one conduction mechanism, which applies to all rocks, the one with $E_a$= 1 eV suggestive of positive hole conductions. If this is true, the different rocks displayed in **Figure 8** would just have different number densities of $h^\bullet$ charge carriers $n_i$ of charge z and mobility μ:

$$\sigma = n_i \, z \, \mu \, \exp [-1 \text{ eV}/kT] \qquad [12]$$

The MgO single crystal studies lead us to the conclusion that, in the temperature range of greatest interest to Earth's crust, where the conditions fall outside thermodynamic equilibrium and rocks are in a metastable state, the *in situ* conductivity is dominated by $h^\bullet$ charge carriers. This points to a near-universal role of peroxy defects and $h^\bullet$ charge carriers in controlling the electrical conductivity of mafic and ultramafic rocks in the upper and middle crust, possible of all rocks in any part of the Earth's crust, where the conditions are ruled by thermodynamic metastability [*Freund*, 2003].

At the same time, if the electrical conductivity of a typical upper mantle rock like dunite as depicted in **Figure 7** appears to be also controlled by $h^\bullet$ charge carriers, this must not be construed to suggest that *in situ*, under upper mantle conditions, olivine contains peroxy defects. This cannot be the case. With reference to **Figure** 1, it can be firmly stated that the temperatures in the upper mantle will be above $T_{freez'g}$, meaning that, under *in situ* conditions, olivine forms stable solid solutions, ss, with the traces of $H_2O$ fluid phase component, which had been incorporated into the crystal matrix at $T_{cryst}$.

However, any rock from high temperature environments collected at the Earth's surface obviously transitioned from the ss field into the sss field and cooled through the 500°C window where the redox conversion exemplified by eqs. [4a/b] takes place. In the case of olivine this conversion will involve hydroxyl pairs of the type $O_3Si\text{-}OH$ splitting off $H_2$ and rearranging under metastable conditions to [*Freund and Oberheuser*, 1986]:

$$O_3Si\text{-}OH\ HO\text{-}SiO_3 \Leftrightarrow O_3Si\text{-}OO\text{-}SiO_3 + H_2 \qquad [13]$$

If the diffusively mobile $H_2$ molecule leaves the site next to the peroxy defect [*Freund et al.*, 2002], the reaction becomes irreversible. Upper mantle olivine crystals collected at the Earth's surface will therefore contain peroxy defects, which – upon reheating in laboratory experiments – release $h^*$ charge carriers:

$$O_3Si\text{-}OO\text{-}SiO_3 + O^{2-} \Leftrightarrow O_3Si\text{-}O{\bullet}O\text{-}SiO_3 + O^- \qquad [14]$$

where $O^-$ stands for $h^\bullet$.

## Conclusion

Peroxy defects consist of pairs of oxygen anions in the valence 1–. Because $O^-$ is more oxidized than $O^{2-}$, peroxy defects are not supposed to exist in minerals and rocks, which, without exception, come from reduced to highly reduced environments. In fact, under thermodynamic equilibrium conditions, peroxy defects are the hallmark of oxidizing conditions, which are not found anywhere deep in the Earth's crust and even less so in the upper mantle. Therefore, finding that common crustal rocks and even upper mantle rocks contain peroxy defects appears to be in violation of thermodynamics. This, however, is not the case.

To study how $O^-$ can be introduced, we present the case of MgO crystals, grown under most reducing conditions in a carbon arc furnace. These MgO crystals contain $OH^-$, forming a solid solution (ss), $Mg_{1-\delta}(OH)_{2\delta}O_{1-2\delta}$ with $\delta \ll 1$, due to small amount of $H_2O$ incorporated during crystallization from the $H_2O$-laden melt. During cooling, all solid state diffusion, specifically diffusion of cations and anions as well as of cation and anion vacancies, slows down, eventually reaching a temperature $T_{freez'g}$, at which the system can be considered frozen. At this point, the systems drifts out of thermodynamic equilibrium and the $Mg_{1-\delta}(OH)_{2\delta}O_{1-2\delta}$ solid solution, ss, turns into a supersaturated solid solution, sss.

While it is true that this transition from equilibrium to non-equilibrium conditions occurs, there is a widespread misconception in the geoscience community that $T_{freez'g}$ marks the point at which all solid state diffusional processes come to a stop.

This is not the case as presented in this paper. Using infrared spectroscopy, it can be shown that, while cooling through a temperature window around 500°C, $OH^-$ pairs at $Mg^{2+}$ vacancy sites in the $Mg_{1-\delta}(OH)_{2\delta}O_{1-2\delta}$ matrix undergo an electronic rearrangement known as a redox conversion (RC). During the RC two $OH^-$ change into a peroxy anion, $O_2^{2-}$, plus an $H_2$. The introduction of peroxy anions into the matrix of the MgO crystals has a profound effect on the electrical conductivity. Using the repetitive heating-cooling-reheating cycles as depicted in **Figures 5a/c**, it can be shown that the break-up of the peroxy anions leads to unbound $O^-$ states, equivalent to defect electrons in the oxygen anion sublattice $h^\bullet$. These $h^\bullet$ are highly mobile electronic charge carriers, associated with the O 2sp-type energy levels at the edge of the valence band, known as positive holes.

In fact, the conductivity of melt-grown MgO single crystals of the highest purity grades increases by 6-7 orders of magnitude during the progressive break-up of the peroxy bonds. Recombination of the thermally activated $h^\bullet$ takes hours to complete at ambient temperature due to their long lifetimes[†]. Therefore, during cooling, many positive holes do not deactivate at the same rate as the temperature decreases. Instead they continue to remain mobile, conducting with the 1 eV activation energy.

The combination of an anomalous conductivity increase during heating and the 1 eV activation energy during cooling creates a diagnostically distinct conductivity response, which a variety of igneous rocks share with MgO. This is a strong indication, if not proof, that the peroxy defects and positive hole charge carriers are not limited to the $MgO–H_2O$ system. The same type of process, the redox conversion to peroxy defects plus $H_2$ molecules, seems to also take place in minerals of igneous rocks. They too contain solute $OH^-$ due to the fact that they crystallized from $H_2O$-laden magmas. During cooling, their solute $OH^-$ pairs obviously also undergo the electronic rearrangement of a redox conversion, producing peroxy defects plus $H_2$ molecules.

On the basis of the work presented here we can say with a high degree of confidence that igneous and high-grade metamorphic rocks in the shallow crust and mid-crustal, ubiquitously contain peroxy defects despite their overall reduced to highly reduced appearance. The peroxy defects will release positive hole charge carriers, for instance during heating, which control the *in situ* electrical conductivity [*Freund*, 2003].

Conversely, minerals and rocks in the lower crust and in the upper mantle cannot contain peroxy defects. The reason is that the temperatures are high enough to activate cation and anion diffusion as well as cation/anion vacancy diffusion. Hence, thermodynamic equilibrium prevails, and there is no possibility for metastable peroxy defects, nor positive hole charge carriers, to exist.

[†] In Part II it will be shown that $h^\bullet$ activated by stress at ambient temperatures appear to have a very wide spectrum of lifetimes, many as short as milliseconds, others as long as minutes to hours, even days to weeks.

In Part II of this two-part paper we'll provide evidence that stresses also activate peroxy defects, generating positive hole charge carriers which can flow out of the stressed rock volume. During propagation through the rock column and upon arrival at the Earth's surface the positive holes cause a number of follow-on processes, from which information can be derived about the build-up of tectonic stresses deep below and, hence, about pre-earthquake signals originating at the Earth's surface that would indicate an increased earthquake risk lurking deep below.

A corollary of this statement is that $h^{\bullet}$ charge carriers can be stress-activated only in portions of the Earth's crust, where metastable conditions prevail and where peroxy defects can generate positive hole charge carriers when challenged by stress. If stress activation of positive holes is a prerequisite for pre-earthquake phenomena, the hypocenters for earthquakes that are preceded by pre-earthquake signals must lie in the upper to middle crust. This conclusion is consistent with reports [*Chen et al.*, 2004; *Liu et al.*, 2006; *Saroso et al.*, 2009] that, with the exception of events in the cold inner part of a subducting plate, ionospheric perturbations are observed primarily before earthquakes with hypocenters in the upper to middle crust.

## Acknowledgments

The results reported in this paper evolved over many years, starting with early studies by Reinhard Martens, Heinz Wengeler and Hendrik Kathrein, supported in part by the Deutsche Forschungs Gemeinschaft, and continuing with work supported at several occasions by the NASA Ames Research Center Director's Discretionary Fund and, most recently, by the NASA Earth Surface and Interior (ESI) program under grant # NNX12AL71G.